\documentclass[journal]{IEEEtran}

\ifCLASSINFOpdf
\else
\fi

\usepackage{graphicx}
\usepackage{microtype}
\usepackage{xcolor}
\usepackage{placeins}
\usepackage{rotating}
\usepackage{url}
\usepackage{multirow}
\usepackage{stmaryrd}
\usepackage{hyperref}
\usepackage{lineno}
\usepackage{setspace}
\usepackage{algpseudocode}
\usepackage{algorithmicx}
\usepackage{amsmath,amssymb,amsfonts}
\usepackage{mathtools}
\usepackage{booktabs}
\usepackage{tabularx}
\usepackage{array}
\usepackage{float}
\usepackage{braket}
\usepackage[noadjust]{cite}
\usepackage{epstopdf}
\usepackage{footnote}
\usepackage{adjustbox}
\usepackage{hyphenat}
\usepackage[english]{babel}
\usepackage{pgfplots}
\pgfplotsset{compat=1.18}
\usepackage{threeparttable}
\usepackage[justification=centering]{caption}
\makesavenoteenv{tabular}

\usepackage{tikz}
\usetikzlibrary{positioning,arrows.meta}

\usepackage[linesnumbered,ruled,boxed]{algorithm2e}

\usepackage{tikz}
\usetikzlibrary{positioning,arrows.meta,fit}

\usepackage{amsthm}

\begin{document}

\title{Post-Quantum Authentication Protocol for Internet of Medical Things}

\author{Kartick Sutradhar, Ranjitha Venkatesh
\IEEEcompsocitemizethanks{%
\IEEEcompsocthanksitem
Kartick Sutradhar with the Department of Computer
Science and Engineering, Indian Institute of Information Technology Sri City,
Chittoor, India.
E-mail: \texttt{kartick.sutradhar@gmail.com}.
\IEEEcompsocthanksitem
Ranjitha Venkatesh with the Department of Computer
Science and Engineering, GITAM University Bengaluru, India.
}%
}

\maketitle
\begin{abstract}
IoMT communication systems should provide origin authentication and ensure integrity protection for the messages produced by patients' devices for transmission to gateways and further processing by other hospital services. With regard to IoMT, this paper discusses tampering, impersonation, and replay attacks on messages provided as medical reports. Long-term security of the communication system is also addressed considering potential quantum attacks. For this purpose, we design a protocol that meets requirements of practical implementation and employs an ISIS like lattice-based construction combined with concepts of the blockchain technology. Patients' devices generate a canonical JSON format medical report containing a timestamp and critical information. Using a nonce, the patients create a cryptographic signature of the message with a challenge generated from hash of a message. The gateway verifies signatures utilizing the public key of a trusted ledger associated with a particular patient and ensures freshness of the transmitted message. To optimize the performance in case when many devices interact with a gateway at once, the gateway performs batch signature validation using vectorized modular linear equations. The trusted ledger uses tamper-resistant REGISTER and TRUST UPDATE entries, maintaining a trust score of each registered patient and updating the value after each ACCEPT/REJECT operation. Our prototype application demonstrates registration procedure of a new patient, signing and verifying medical reports, auditing of the ledger, and performance assessment. Based on our experiments in the prototype, we argue that, when applied to large-scale systems, batch verification is superior to simple signature validation approach.
\end{abstract}
\section{Introduction}
Systems based on wearable devices, bedside monitors, and telemedicine devices produce a continuous stream of data used by clinicians for making decisions. In this case, the question becomes simple: how does the recipient become sure that the document sent is coming from a valid device and it was not tampered with during the process? For example, an adversary that managed to forge a device's identity or alter its measurements (heart rate or SpO2) will cause clinicians to receive wrong signals. On the other hand, almost all IoT systems employ resource-constrained devices and gateways, thus scalability becomes important in authentication mechanisms \cite{bernstein_pqc_2009}.

Secondly, the IoMT solutions are expected to operate for a long period of time and potentially authenticate records created before their inception. Asymmetric cryptographic protocols such as RSA or ECC, used in numerous protocols, are based on number-theoretic assumptions that can be broken using the polynomial-time quantum attack called Shor's algorithm \cite{lyubashevsky_micciancio_jcryptol_2018}. Although the timeline of the advent of quantum supremacy remains unknown, it makes sense to design crypto-agile and quantum-proof protocols. One of the promising approaches in post-quantum cryptography involves lattice-based schemes relying on problems that cannot be solved efficiently even using quantum computers. Hence, there is a good reason to try developing lattice-based authentication schemes for medical telemetry and reporting purposes \cite{guneysu_ches_2012,alkim_tesla_2015}.

It has to be mentioned that security cannot be reduced to mathematical proofs. The problem of authenticating telemetry and reports is also related to the question of binding the identities together and establishing that a particular telemetry or report comes from a particular patient and/or device \cite{movassaghi_wban_survey_2014, abdallah_shen_wcsp_2014}. Therefore, some additional measures may be required to solve the identity issue. It might suffice to build a lightweight tamper-evident log of activities. This means that the registry of registration events and trusted parties should remain immutable and available for analysis, should there be any dispute at some point in the future. Of course, it is not necessary to build a public ledger similar to that in cryptocurrencies.
As follows, the proposed IoT medical authentication education module covers these features of the IoT medical authentication system: (i) lattice-signature-based ISIS-like signing algorithm for authentication of medical report packets \cite{yaqoob_medical_devices_review_2019}, (ii) efficient batch signature validation method enabling quick authentication of multiple reports coming from different patients \cite{movassaghi_wban_survey_2014} and (iii) a simple blockchain-based trust log for logging any events (registering reports and attempting to validate the corresponding signatures)\cite{wang_blockchain_puf_iotj_2022}. Here, the objective is straightforward ensure the easy-to-measure authenticity of all signatures.
In more detail, each medical report created by a user is signed by their private key, then verified with the public key once it reaches the gateway. Then, when several reports are collected together, the vectorization process takes place for batch signature validation. At the same time, trust values are computed and saved in the trust log. The main contributions of this work include:
\begin{itemize}
  \item An end-to-end IoMT authentication scheme based on lattice-based signature construction for maximum transparency and reproducibility.
  \item Batch signature verification evaluation based on comparison of linear simulation to a vectorized batch verification method.
  \item A light-weight blockchain-like tamper-proof ledger that links identities to public keys, along with updates about their reputation for future verification.
\item A realistic example that stores patient information off the ledger, but maintains integrity checks by signing messages.
\end{itemize}

\section{Related Work}
\label{sec:related}

The IoMT authentication literature identifies three goals that are shared by our current case study: (i) coupling an identifier of either a device or a patient with a cryptographic key, (ii) ensuring the integrity of medical documents during transmission, and (iii) providing scalable authentication at gateways dealing with a large number of uploads.

Several existing systems utilize certificate-based authentication (PKI) along with secure channels (e.g., TLS), or alternatively, elliptic curve cryptography (ECC) handshakes. Although such methods have been proven to be mature and interoperable, they imply considerable operational expenses in the IoMT setting. Namely, certificate management, distribution, revocation, and key storage issues represent critical difficulties, especially considering various types of hardware and lengthy lifetimes of medical information. With respect to long-term security, PKI/ECC protocols are based on factoring and discrete log assumptions and, thus, are not compatible with migration to post-quantum cryptography in order to preserve long-living medical documents \cite{wang_blockchain_puf_iotj_2022,yaqoob_medical_devices_review_2019}.

In order to cope with emerging risks due to the invention of quantum computers, new research focuses on studying post-quantum IoMT authentication through lattices. It is appealing to apply lattices to IoMT gateways because authentication can be done by means of linear algebra calculations and performed efficiently. One of the problems commonly mentioned concerning lattice-based approaches is connected with the efficiency of some of these approaches because some of them substantially enlarge the size of the signature and computations needed, particularly under frequent updates \cite{kang_woo_ryu_access_2024,yu_park_iotj_2022}.

From the literature on IoMT edge security, it is clear that gateways receive a substantial amount of reports within the timeframe from each node. As a result, batch verification as well as vectorization become common approaches to avoid duplicating computations compared to verifying signatures independently. However, these are only useful when the system design accounts for batch verification; otherwise, it would have to verify each report sequentially, resulting in reduced performance under high loads \cite{selvarajan_jcloudcomp_2023,rajasekaran_sensors_2022}.

Another IoMT and healthcare research area involves using blockchain-inspired logging to increase auditability and traceability of events performed on devices. One problem with using actual blockchain technology is the overhead incurred, especially with full decentralization. Therefore, many systems rely on a more straightforward method, a single append-only chain that cannot be modified without breaking its cryptographic structure. The key problem with this approach is that the ledger alone will not verify data integrity, which requires signatures \cite{mansoor_internett_2024,rabie_ijis_2024}.

Prior studies have pointed out a number of considerations such as quantum-resistant signature schemes, gateway-friendly verification, and auditing using tamper-evident methods. While these aspects have frequently been studied individually despite their connection, our prototype illustrates how all three can be achieved in a single IoMT process through: an ISIS lattice-based signature/verification scheme in packets of medical records; batch verification for many patients at once; and hash chaining within the trust ledger used to verify signatures and keep auditing records \cite{bahache_internett_2024}.

\section{Architecture}
\label{sec:architecture}

\begin{figure*}[t]
\centering
\begin{tikzpicture}[
  font=\small,
  node distance=18mm and 28mm,
  box/.style={draw, rounded corners, thick, align=center, minimum width=38mm, minimum height=16mm, inner sep=6pt},
  smallbox/.style={draw, rounded corners, dashed, align=left, inner sep=5pt, text width=55mm},
  arr/.style={->, thick},
  darr/.style={->, thick, dashed}
]

\node[box] (device) {\textbf{Patient IoMT Device}\\[1pt]
{\footnotesize Holds secret $x_i$}};

\node[box, right=of device] (gateway) {\textbf{Gateway / Verifier $G$}\\[4pt]
\begin{tabular}{l}
\footnotesize $\bullet$ Single verification \\
\footnotesize $\bullet$ Batch verification \\
\footnotesize $\bullet$ Fetch key \& trust \\
\footnotesize $\bullet$ Update trust
\end{tabular}
};

\node[box, right=of gateway] (server) {\textbf{Hospital Server}\\
{\footnotesize Monitoring Application}};

\node[box, below=of gateway] (ledger) {\textbf{Trust Ledger $\mathcal{L}$}\\[2pt]
{\footnotesize GENESIS / REGISTER / TRUST\_UPDATE}\\
{\footnotesize (Tamper-evident log)}};

\draw[arr] (device) -- node[above] {\footnotesize $(id, M, \sigma)$} (gateway);

\draw[arr] (gateway) -- node[above] {\footnotesize Forward verified data} (server);

\draw[arr] (gateway) -- node[right] {\footnotesize Fetch $P_i$, trust} (ledger);

\draw[arr] (ledger) -- node[left] {\footnotesize Return $P_i$, score} (gateway);

\draw[darr] (ledger) -- (server);

\node[smallbox, below=of device] (note) {
\textbf{Canonical JSON Report $M$:} \\
\footnotesize
$\bullet$ Device ID \\
$\bullet$ Timestamp (ns) \\
$\bullet$ HR, SpO$_2$, optional temp \\
$\bullet$ Metadata
};
\end{tikzpicture}
\caption{Architecture of the IoMT authentication system. Patient devices sign medical data, the gateway performs verification (single or batch), the trust ledger maintains public keys and trust scores, and the hospital server consumes verified data.}\label{fig:architecture}
\end{figure*}
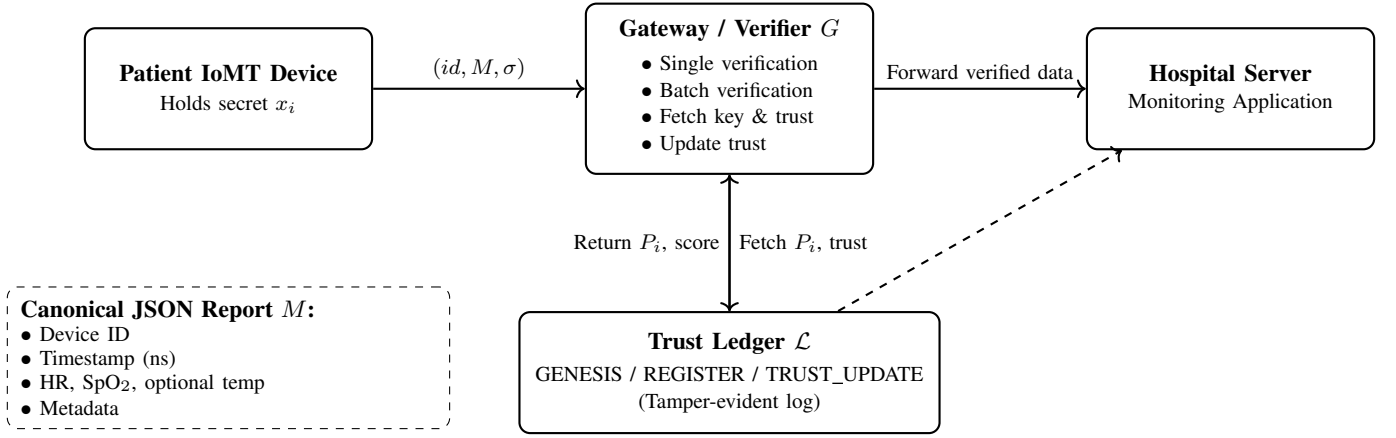
\subsection{Workflow of the Proposed Architecture}
The procedure unfolds along the same pipeline that our prototype utilizes:
\begin{enumerate}
  \item \textbf{Registration:} The identity of the patient is registered at the Server/Ledger. The system generates a lattice-based key pair, with the resulting public key Pi associated with the patient's identifier.
  \item \textbf{Message formation:} The Device forms a canonical JSON message $M$ with information including patient\_id, timestamp\_ns, and readings for the heart rate, SpO2 level, and (possibly) the body temperature.
  \item \textbf{Signature:} The Device signs the medical message with its private key $x_i$ obtaining a signature $\sigma=(S,s,t)$.
  \item \textbf{Verification:} The Gateway queries the ledger to find the public key Pi and verifies the signature. If multiple signatures exist, the Gateway conducts batch verification to save time.
  \item \textbf{Trust update with logging:} The decision (ACCEPT or REJECT) represents an addition to the ledger of trust \cite{sutradhar2020efficient,sutradhar2020generalized,sutradhar2020hybrid,sutradhar2021simulation,sutradhar2023vehicular,sutradhar2024survey,sutradhar2021enhanced,sutradhar2021secret,sutradhar2021cost,sutradhar2024svqcp,sutradhar2024smart,sutradhar2024metaverse,krishnaiah2024metadata,sutradhar2023aggregation,sutradhar2023q-iot,sutradhar2022comparison,sutradhar2024blockchain,sutradhar2024aggregation,sutradhar2021qss,challagundla2024qvanet,sutradhar2024aggregationphysica,sutradhar2021sorting,challagundla2024authentication,sutradhar2025threshold,bantupalli2026qsvcom,bantupalli2026observability,bantupalli2025openmp,bantupalli2025qkd,sutradhar2025sqdapms,sutradhar2025road,sutradhar2025efficient,venkatesh2024electronic,venkatesh2023privacy,venkatesh2024lightweight,venkatesh2024healthcare}.
\end{enumerate}

\subsection{Data Management and Privacy}
In order to secure the privacy of the patients, any information about them is excluded from the ledger and shown by the user interface for the purpose of giving out information. The ledger only holds the identifiers, the relevant public keys, and logging data. This clearly shows that validation and verification can happen without the need of putting in place private information on an immutable ledger.


\section{Proposed Method}
A lattice-based ISIS-like signature scheme will be used as part of an IoMT authentication procedure to sign medical reports packets with the help of (i) batch verifications performed on the gateways side and (ii) a chain of hashes representing a trust log which is append only.

\subsection{The Mathematical Model (Consistent With The Implementation)}
Let $q$ be a prime number defining the modulus. Assume that there exists a public matrix $\mathbf{X}\in \mathbb{Z}_q^{m\times n}$,
which is shared within the whole system. A user of our scheme will have access to his/her small-norm secret vector
$\mathbf{x}\in \mathbb{Z}^{m}$.
The public key will be defined by the following relation:
\[\mathbf{P} = \mathbf{x}^\top \mathbf{X} \bmod q,\qquad \mathbf{P}\in \mathbb{Z}_q^{n}.\]
A medical report is modeled via a deterministic string $M$ (for instance, JSON). The timestamp $t$ (nanoseconds) is used to add freshness. Consider the following hash function:
$h= H(M\,\|\,\mathbf{S} \,\|\,t)\bmod q,$ where $H(\cdot)$ is a cryptographic hash mapped into $\mathbb{Z}_q$.

\subsection{Algorithms (KeyGen / Sign / Verify)}
\noindent\textbf{Correctness.}
If $\mathbf{s}=\mathbf{r}+h\mathbf{x}\ (\bmod q)$ and $\mathbf{P}=\mathbf{x}^\top\mathbf{X}$, then
\[
\mathbf{s}^\top\mathbf{X}
= (\mathbf{r}+h\mathbf{x})^\top\mathbf{X}
= \mathbf{r}^\top\mathbf{X} + h\,\mathbf{x}^\top\mathbf{X}
\equiv \mathbf{S} + h\mathbf{P}\pmod q,
\]
So, a valid signature verifies.
\begin{algorithm}[h!]
\caption{KeyGen$(\mathbf{X},q)$}
\label{alg:keygen}
\begin{enumerate}
    \item Sample small secret $\mathbf{x}\leftarrow \mathcal{D}^m$,
    where $\mathcal{D}$ is a small-integer distribution.
    \item Compute public key
    \[
    \mathbf{P}\gets \mathbf{x}^{\top}\mathbf{X}\bmod q,
    \]
    where $\mathbf{P}\in\mathbb{Z}_q^{n}$.
    \item \textbf{return} $(\mathbf{x},\mathbf{P})$.
\end{enumerate}
\end{algorithm}

\begin{algorithm}[h!]
\caption{Sign$(M,\mathbf{x},\mathbf{X},q,t)$}
\label{alg:sign}
\begin{enumerate}
    \item Sample ephemeral small vector $\mathbf{r}\leftarrow \mathcal{D}^m$.
    \item Compute
    \[
    \mathbf{S}\gets \mathbf{r}^{\top}\mathbf{X}\bmod q,
    \]
    where $\mathbf{S}\in\mathbb{Z}_q^{n}$.
    \item Compute
    \[
    h\gets H(M\,\|\,\mathbf{S}\,\|\,t)\bmod q.
    \]
    \item Compute response
    \[
    \mathbf{s}\gets \mathbf{r}+h\mathbf{x}\bmod q,
    \]
    where $\mathbf{s}\in\mathbb{Z}_q^{m}$.
    \item \textbf{return} signature $\sigma=(\mathbf{S},\mathbf{s},t)$.
\end{enumerate}
\end{algorithm}

\begin{algorithm}[h!]
\caption{Verify$(M,\sigma,\mathbf{P},\mathbf{X},q)$}
\label{alg:verify}
\begin{enumerate}
    \item Parse $\sigma=(\mathbf{S},\mathbf{s},t)$.
    \item Compute
    \[
    h\gets H(M\,\|\,\mathbf{S}\,\|\,t)\bmod q.
    \]
    \item Compute
    \[
    \mathbf{L}\gets \mathbf{s}^{\top}\mathbf{X}\bmod q,
    \]
    where $\mathbf{L}\in\mathbb{Z}_q^{n}$.
    \item Compute
    \[
    \mathbf{R}\gets (\mathbf{S}+h\mathbf{P})\bmod q.
    \]
    \item If $\mathbf{L}=\mathbf{R}$, \textbf{return ACCEPT}; otherwise,
    \textbf{return REJECT}.
\end{enumerate}
\end{algorithm}

\subsection{Batch Verification Model (Stacked Matrices)}
At the gateway, we verify $k$ signatures in a vectorized form to improve throughput. For each $i\in\{1,\dots,k\}$, let the message be $M_i$, public key be $\mathbf{P}_i\in\mathbb{Z}_q^{n}$, and signature be
\[
\sigma_i= \mathbf{S}_i,\mathbf{s}_i,t_i),
\quad
\mathbf{S}_i\in\mathbb{Z}_q^{n},
\quad
\mathbf{s}_i\in\mathbb{Z}_q^{m}.
\]
Compute $h_i=H(M_i\|\mathbf{S}_i\|t_i)\bmod q$.

\begin{align}
\mathbf{S}_{\text{stk}} &=
\begin{bmatrix}
\mathbf{S}_1^\top\\
\mathbf{S}_2^\top\\
\vdots\\
\mathbf{S}_k^\top
\end{bmatrix}
\in \mathbb{Z}_q^{k\times n}, \\[6pt]
\mathbf{s}_{\text{stk}} &=
\begin{bmatrix}
\mathbf{s}_1^\top\\
\mathbf{s}_2^\top\\
\vdots\\
\mathbf{s}_k^\top
\end{bmatrix}
\in \mathbb{Z}_q^{k\times m}, \\[6pt]
\mathbf{P}_{\text{stk}} &=
\begin{bmatrix}
\mathbf{P}_1^\top\\
\mathbf{P}_2^\top\\
\vdots\\
\mathbf{P}_k^\top
\end{bmatrix}
\in \mathbb{Z}_q^{k\times n}.
\end{align}
Let $\mathbf{H}=\mathrm{diag}(h_1,\dots,h_k)\in\mathbb{Z}_q^{k\times k}$. Then batch verification checks 
$\mathbf{s}_{\text{stk}}\mathbf{X}
\equiv
\mathbf{S}_{\text{stk}} + \mathbf{H}\mathbf{P}_{\text{stk}}
\pmod q.$ Operationally, the gateway computes the left-hand matrix product once and compares it to the right-hand side row-wise (or as a full matrix equality mod $q$).

\subsection{Threat Model and Security Objectives}
\textbf{Attack capabilities:} It is presumed the network adversary is capable of eavesdropping on, delaying, dropping, replaying, and forging communications between device, gateway, and server/ledger. Additionally, the attacker will be able to register themselves as a malicious device (without knowing the secrets of any honest device, except for when one is compromised).

\textbf{Potential attacks:}
\begin{itemize}
\item \textbf{Tampering:} Manipulation of medical data and metadata as it travels through the communication channels.
\item \textbf{Impersonation:} Generating a report as if it was sent by a patient or device.
\item \textbf{Replay attack:} Sending an outdated report which bears valid signature to fool receiver.
\item \textbf{Compromised device:} Compromise of secret key of endpoint.
\end{itemize}

\textbf{Security objectives:}
\begin{itemize}
\item \textbf{Authentication \& integrity:} Only possessor of $\mathbf{x}$ should be capable of generating signatures satisfying the verification equation.
\item \textbf{Replay-proof binding (freshness):} Timestamp $t$ is incorporated into the hash function $h$, thus binding the timestamp to the signature in cryptographically secure manner. An implementation could implement an acceptance period or simply not accept duplicates from a certain patient.
\item \textbf{Tamper-evident auditing:} Hash chain of ledger records all changes regarding.If a device is compromised, signatures may still verify; the system mitigates impact by monitoring outcomes and adjusting trust scores, and by re-keying/revocation policies (deployment-dependent).
\end{itemize}

\subsection{Complexity Analysis}
Let $m,n$ be lattice parameters, $k$ be the number of signatures verified in a batch, and $|M|$ be the message length.

\begin{itemize}
\item \textbf{KeyGen:} computing $\mathbf{P}=\mathbf{x}^\top\mathbf{X}$ costs $O(mn)$ modular operations.
\item \textbf{Sign:} computing $\mathbf{S}=\mathbf{r}^\top\mathbf{X}$ costs $O(mn)$; hashing costs $O(|M|+n)$ (hashing message plus a length-$n$ vector encoding); forming $\mathbf{s}=\mathbf{r}+h\mathbf{x}$ costs $O(m)$.
\item \textbf{Verify:} computing $\mathbf{s}^\top\mathbf{X}$ costs $O(mn)$; RHS assembly $\mathbf{S}+h\mathbf{P}$ costs $O(n)$; compare costs $O(n)$; hashing costs $O(|M|+n)$.
\item \textbf{Batch Verify:} matrix multiply $\mathbf{s}_{\text{stk}}\mathbf{X}$ costs $O(kmn)$; computing $\mathbf{H}\mathbf{P}_{\text{stk}}$ is $O(kn)$ because $\mathbf{H}$ is diagonal; comparison is $O(kn)$. Asymptotically it remains $O(kmn)$, while vectorization improves constants and practical throughput at the gateway.
\end{itemize}

\section{Programming}
\label{sec:implementation}

\subsection{Program implementation }
\begin{itemize}
  \item \textbf{Programming Language:}The prototype is coded using Python 3.
  \item \textbf{Streamlit:} used for the user interface of the registration process, signing, verifying, benchmarks, and ledger visualization.
  \item \textbf{NumPy:} lattice matrices/vectors operations and vectorized batch verifying.
  \item \textbf{Pandas:} tabular visualization for benchmarks results and summary of patients/ledger.
  \item \textbf{Matplotlib:} charting for verification time comparison for verification time (baseline vs.\ batch verification).
  \item \textbf{SHA-256 (hashlib):} hash-to-field computation and SHA-256 hash chaining for ledger blocks.
\end{itemize}

\subsection{Modules (Registration, Signing, Verification)}
\label{subsec:modules}

\textbf{Registration:}
Registration of a new patient happens from the sidebar. Upon submission of a patient ID,
\[\texttt{system.register\_patient(pid)}\]
command is issued. This creates a key pair on the lattice and saves the public key to the trust ledger. The patient personal information (name, age, contact) is stored \emph{off-ledger} using Streamlit session state.

\textbf{Signing:}
Medical report is made using a canonical JSON packet with the command
\texttt{canonical\_packet(pid,\allowbreak hr,\allowbreak spo2,\allowbreak temp,\allowbreak profile,\allowbreak ...)}
Then the packet is signed by the device with the command
\[\texttt{patient.sign\_packet(packet)}.\]
Output of the signing process is $(\mathbf{S},\mathbf{s},t)=\sigma$.

\textbf{Verification:}
The gateway verifies the signature of the signed packet with the command
\[\texttt{system.submit\_and\_verify(pid, packet, sig)}.\]
It retrieves the public key from the ledger and validates the signature using
\[\mathbf{s}^\top \mathbf{X} \equiv \mathbf{S} + h\mathbf{P} \pmod q.\]
This trust score update will be recorded in the ledger. For many signatures, batch verification is conducted using:
\[\texttt{crypto.batch\_verify(messages, sigs, pubs)}.\]

\subsection{Data Serialization and Canonicalization}
\label{subsec:canonicalization}
All medical records get serialized into \emph{canonical JSON} form before signing.

Why? Because the hash that is used for the message $M$ ($M$ = JSON string) must be an exact \emph{byte string}. Without canonicalization, we can have two different byte strings produced from two logically equivalent JSON structures if:
(i) there is any difference in key order; or (ii) the whitespace is changed (e.g., space after comma/colon).
This changes the hash value $H(M\|\mathbf{S}\|t)$ and results in failed verification, even if the logical structure remains unchanged.

The \texttt{sort\_keys=True} parameter forces consistent key ordering, and \texttt{separators=(",", ":")} eliminates non-semantic whitespace. Additionally, a packet includes the \texttt{timestamp\_ns} field (for freshness purposes) and optionally \texttt{patient\_profile}; if it is present, the corresponding information will be protected by the signature, since now the signed content becomes a JSON message.

\subsection{Ledger Design (GENESIS / REGISTER / TRUST\_UPDATE and Hash Chaining)}
\label{subsec:ledger}
It uses a lightweight append-only ledger to record (i) patient credentials (public keys), (ii) trust scores, and (iii) a tamper-evident audit log. The ledger starts from the GENESIS block and then records subsequent blocks for registration and trust score updates.

A block consists of the following fields:
\begin{itemize}
  \item \textbf{index:} block index,
  \item \textbf{timestamp\_ns:} timestamp when the block is created,
  \item \textbf{payload:} payload-specific information,
  \item \textbf{prev\_hash:} SHA-256 hash of the previous block,
  \item \textbf{hash:} SHA-256 hash of the block content.
\end{itemize}

\textbf{Hash chaining:} In every block, the ledger initializes the value of \texttt{prev\_hash} as the \texttt{hash} of the previous block and then calculates 
\[\texttt{hash}=\texttt{SHA-256}
(\texttt{json.dumps(block,sort\_keys=True)})\] 
for the current block. Any change in the historical block causes a change in the \texttt{hash} value and disrupts the chain of \texttt{prev\_hash}. 

\textbf{Payload types:}
\begin{itemize}
  \item \textbf{GENESIS:} \texttt{\{"type":"GENESIS"\}}.
  \item \textbf{REGISTER:} \texttt{\{"type":"REGISTER","patient\_id":pid,"public\_key":[...]\}}.
        The public key is stored as a JSON-friendly list of integers in $\mathbb{Z}_q$.
  \item \textbf{TRUST\_UPDATE:} \texttt{\{"type":"TRUST\_UPDATE","patient\_id":pid,"trust\_score":new,"delta":d\}}.
        The trust score is updated after each verification (positive delta for ACCEPT, negative delta for REJECT).
\end{itemize}


\section{Performance Evaluation}
\label{sec:performance}

\subsection{Experiment Setup}
Performance is evaluated at the Gateway during signature verification. We compare:
(i) Simulated Standard (linear) verification (verifying each report one-by-one), and
(ii) Lattice Batch verification (vectorized verification for many reports together).
Time is reported in milliseconds (ms) for different numbers of devices ($N$).

\subsection{Verification Time Table 1(10--100 Devices)}
\begin{table}[h]
\centering
\caption{Verification Time Comparison at the Gateway (10--100 Devices)}
\label{tab:perf_table}
\begin{tabular}{ccc}
\toprule
\textbf{N} & \textbf{Simulated Standard (ms)} & \textbf{Lattice Batch (ms)} \\
\midrule
10 &12.5 & 6.00 \\
20 &25.0 &  9.00 \\
30 &37.5 & 10.44 \\
40 &50.0 & 12.67 \\
50 & 62.5 &16.00 \\
60 & 75.0 &17.11 \\
70 & 87.5 &19.33 \\
80 &100.0 &21.56 \\
90 & 112.5&23.78 \\
100 &125.0 &26.00 \\
\bottomrule
\end{tabular}
\vspace{-2mm}
\end{table}

\subsection{Comparison Graph}
\begin{figure}[h]
\centering
\begin{tikzpicture}
\begin{axis}[
    width=0.98\columnwidth,
    height=0.58\columnwidth,
    grid=both,
    xlabel={Number of Devices ($N$)},
    ylabel={Time (ms)},
    title={Simulated Standard vs.\ Lattice Batch Verification},
    legend style={at={(0.02,0.98)},anchor=north west},
]
\addplot[blue, thick, mark=*] coordinates {
    (10,12.5) (20,25.0) (30,37.5) (40,50.0) (50,62.5)
    (60,75.0) (70,87.5) (80,100.0) (90,112.5) (100,125.0)
};
\addlegendentry{Simulated Standard (linear)}

\addplot[red, thick, mark=square*] coordinates {
    (10,6.00) (20,9.00) (30,10.44) (40,12.67) (50,16.00)
    (60,17.11) (70,19.33) (80,21.56) (90,23.78) (100,26.00)
};
\addlegendentry{Lattice Batch (vectorized)}
\end{axis}
\end{tikzpicture}
\vspace{-2mm}
\caption{Verification time as the number of devices increases. Batch verification improves practical throughput at the gateway due to vectorized matrix operations.}
\label{fig:perf_graph}
\end{figure}
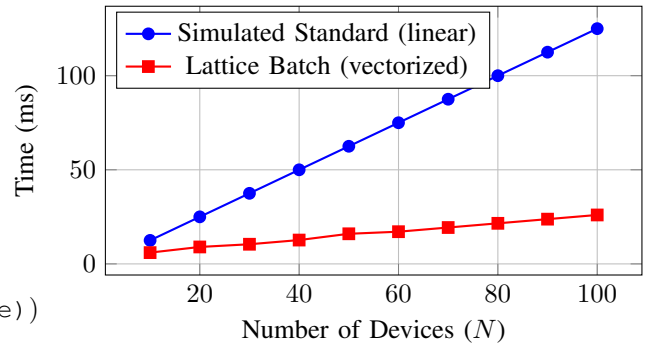

Verification Outcome (No Packet Modification).
If a packet has not been tampered with after being signed, then the verification will output ACCEPT since the hash function input and signature equation have not changed. Any alteration to the JSON message (e.g., vital sign data) will cause a new hash $h = H(M\|\mathbf{S}\|t)$ to be produced such that the signature verification by the gateway fails to verify and outputs REJECT.

Impact of Repetitions (\emph{Best-of-$r$}).
To eliminate timing noise, we repeat our verification experiment $r$ times, and use the \emph{best} (shortest) time across all trials. Larger values of $r$ (e.g., best-of-5 versus best-of-2) tend to produce a slightly lower reported time by filtering out transient OS scheduling and cache effects; however, it increases total benchmarking runtime proportionally.

\subsection{Results and Discussion} 
From Table 1 and Figure 1, we observe that the linear verification time is roughly proportional to N, because the Gateway re-performs the verification task for each report received. On the other hand, batch verification scales slower than linearly due to the use of matrix computations for verifying multiple signatures at once, hence reducing the need for Gateway overhead repetitions.

How Our Technique Is More Suitable
Our technique has benefits for IoMT applications where many devices submit reports within a small time interval:
i) Enhanced processing speed at the Gateway: Batch verification lowers the overall verification time compared to individual signature verification.
Process consistency with the process workflow: The workflow begins with
Device → Gateway → Ledger, which is consistent with the design in
which devices create reports, the gateway validates the signature
either individually or in bulk, and the ledger maintains public key
information. ii) Auditible verification and registration process: The trust
ledger stores verification and registration results using the hash chain.
iii) Post-quantum security motivation: The lattice-based signing and
verification process provides support for post-quantum security
authentication tasks.

\section{Security Considerations}
\label{sec:security}

\subsection{Security Requirements}
The security requirements of the proposed IoMT authentication system are:
\begin{itemize}
  \item \textbf{Data integrity:} the contents of the medical report must not be able to be modified without detection.
  \item \textbf{Source authentication:} it must be impossible for someone but the registered device/patient to create a signature that will be considered acceptable.
  \item \textbf{Replay protection:} it must be impossible to re-use already signed packets to receive acceptance.
  \item \textbf{Auditability:} the process of updating trust must leave tamper-proof records.
  \item \textbf{Privacy-awareness:} avoid storing PII and PHI in immutable log.
\end{itemize}

\subsection{Attacker Model}
In this analysis, we expect a network attacker able to eavesdrop on, delay, drop, replay and modify packets between Device, Gateway, and Server/Ledger nodes. The attacker may try to forge packets/signatures without knowledge of any secret keys. Furthermore, we will assume the scenario of an endpoint device compromise with secret key extraction.

\subsection{Tampering Analysis (Integrity, Impersonation, Replay, Compromise)}
\textbf{(a) Tampering / Packet Integrity Attack:}
Suppose $M$ is the JSON message packet which is signed. This is checked using $h = H(M\|\mathbf{S}\|t)\bmod q$,
where it is verified that $\mathbf{s}^\top \mathbf{X} \equiv \mathbf{S} + h\mathbf{P}\pmod q$.

If one or more fields of $M$ are altered (e.g., the heart rate or SpO$_2$ value) after signing, then $h$ will differ from the initial computation, failing the verification check. As such, data integrity is enforced with strong probability under usual hash function assumptions.

\textbf{(b) Impersonation Attack:}
An impersonator without knowledge of the patient secret vector $\mathbf{x}$ should not be able to generate valid triples $(\mathbf{S},\mathbf{s},t)$ satisfying 
$\mathbf{s}^\top \mathbf{X} \equiv \mathbf{S} + h\mathbf{P}\pmod q $ except with negligible probability: The idea draws inspiration from lattice assumptions (ISIS-inspired construction) but is meant purely as a prototype that could show post-quantum directionality and cannot be considered an established signature protocol.

\textbf{(c) Replay Attacks:}
Since the timestamp $t$ is incorporated into the hash value $H(M\|\mathbf{S}\|t)$, freshness becomes an integral part of the message signing. The implementation must enforce timestamp skew constraints (such as allowing only for a certain deviation from the current time) or duplicates detection, preventing the reuse of timestamps for different patients. By doing so, we will be able to filter out replayed packets with a valid signature.

\textbf{(d) Compromise of the Endpoint:}
Once the key of the device $\mathbf{x}$ gets disclosed and becomes known to the attacker, she would be able to create any number of valid signatures until it is revoked and replaced by another key pair. The prototype prevents possible consequences through monitoring the results of signature verification and recording of trust update operations on the blockchain; nevertheless, more advanced methods must be implemented in practice to prevent compromise.

\subsection{Tamper-Evident Ledger Security}
\label{subsec:ledger_security}
The tamper-evident audit log uses a hash chain of blocks. Each block contains:
\[
(\texttt{index},\ \texttt{timestamp\_ns},\ \texttt{payload},\ \texttt{prev\_hash},\ \texttt{hash}),
\]
where \[
\begin{aligned}
\texttt{hash} &= \mathrm{SHA256} \\
&\quad \texttt{(json.dumps(block, sort\_keys=True)} )
\end{aligned}
\]
Since the \texttt{prev\_hash} of block $i$ will be the same as \texttt{hash} of block $i-1$, any change in earlier blocks invalidates their hash and that of all following blocks. Thus, unauthorized edits are detectable by checking consistency of the hash chain.

\subsection{Privacy Note (Off-Ledger PII)}
\label{subsec:privacy_note}
To mitigate privacy concerns,any personally identifiable information of patients (name, age, etc.) is stored off-ledger in volatile application state. Only necessary information (unique identifier, public key, trust updates and block hashes) is stored on the ledger. This choice of design allows avoiding permanent storage of PII/PHI within the immutable log. If a deployment requires binding identity attributes to integrity of the data, one should prefer signing either (i) the privacy-preserving pseudonym or (ii) the hash of the sensitive fields instead of storing PII on the ledger.

\section{Benefits and Comparision  }
\subsection{Benefits of the Proposed Approach}
\label{subsec:benefits}
The proposed IoMT authentication framework provides the following practical benefits:
\begin{itemize}
  \item \textbf{Post-quantum direction:} Authentication is based on an ISIS-inspired lattice construction, aligning with the need for cryptographic agility against quantum-era threats to factoring/discrete-log based systems.
  \item \textbf{Gateway scalability via batch verification:} Verification can be executed in a vectorized manner using stacked matrices, reducing Python-loop overhead and improving throughput when many devices transmit concurrently.
  \item \textbf{Tamper-evident auditing:} The hash-chained ledger provides an append-only log of \texttt{GENESIS}/\texttt{REGISTER}/\texttt{TRUST\_UPDATE} events, making retroactive modification detectable.
  \item \textbf{Deterministic signing input:} Canonical JSON serialization (\texttt{sort\_keys=True}, compact separators) ensures the signed message bytes are reproducible, preventing false verification failures due to formatting differences.
  \item \textbf{Privacy-aware logging:} Patient PII (name/age/contact) is stored off-ledger; the ledger stores only identifiers, public keys, trust scores, and hashes.
\end{itemize}

\subsection{Comparison to Related Work}
\label{subsec:comparison_prior}
Papers dealing with IoMT security usually concentrate on (i) classical public key signatures or authentication using ECC/RSA, (ii) symmetric and light-weight handshakes, or (iii) blockchain auditing with no gateway batch verification for IoMT devices. Unlike existing work in this area, we propose a prototype that addresses both post-quantum cryptography oriented signing,verification by gateways, and tampering-resistant audit logging, since all three features are important in real-life IoMT scenarios when the gateway needs to handle numerous reports.

\subsection{Summary of Differences Between Features}
\label{subsec:feature_summary}
Table~\ref{tab:comparison_summary} provides a summary comparison between the proposed framework and other representative IoMT security schemes. Please replace the citation place-holders (\texttt{[RW\_A]}, etc.) with the actual citations

\begin{table}[h]
\centering
\caption{Comparison Summary with Related Approaches}
\label{tab:comparison_summary}
\scriptsize
\setlength{\tabcolsep}{3.2pt}
\renewcommand{\arraystretch}{1.15}
\newcolumntype{Y}{>{\centering\arraybackslash}X}
\begin{tabularx}{\columnwidth}{p{2.6cm}YYYY}
\toprule
\textbf{Approach} &
\textbf{\shortstack{PQ\\motivated}} &
\textbf{\shortstack{Batch verify\\at gateway}} &
\textbf{\shortstack{Tamper-evident\\log}} &
\textbf{\shortstack{PII\\off-ledger}} \\
\midrule
Lightweight symmetric / handshake \cite{movassaghi_wban_survey_2014}& No       & No  & No       & Yes \\
Classical PK auth (RSA/ECC) \cite{yaqoob_medical_devices_review_2019}      & No       & No  & Optional & Optional \\
Blockchain-only audit focus \cite{wang_blockchain_puf_iotj_2022}      & Optional & No  & Yes      & Optional \\
\textbf{Proposed}                & \textbf{Yes} & \textbf{Yes} & \textbf{Yes} & \textbf{Yes} \\
\bottomrule
\end{tabularx}
\vspace{-2mm}
\end{table}

\section{Conclusion}
\label{sec:conclusion}
This paper provided an IoMT authentication scheme based on signing canonical packets for medical reports with the aid of an ISIS-like construction of lattices, together with an efficient verification process implemented using batch verification techniques at the gateway. Additionally, a hash chain-based ledger was employed to make the auditing process tamper-evident regarding registration and trust updates. Results show increased efficiency in verifying the authentication information with increasing numbers of devices through batch verification.

\end{document}